\documentstyle{camera}
\newcommand{\AmS}{{\protect\the\textfont2
  A\kern-.1667em\lower.5ex\hbox{M}\kern-.125emS}}

\newcommand{\beq}{\begin{equation}}
\newcommand{\eeq}{\end{equation}}
\newcommand{\bea}{\begin{eqnarray}}
\newcommand{\eea}{\end{eqnarray}}

\def\dm2{\Delta m^2}
\def\sq2{sin^2(2\Theta)}

\def\la{\mathrel{\mathchoice {\vcenter{\offinterlineskip\halign{\hfil
$\displaystyle##$\hfil\cr<\cr\sim\cr}}}
{\vcenter{\offinterlineskip\halign{\hfil$\textstyle##$\hfil\cr<\cr\sim\cr}}}
{\vcenter{\offinterlineskip\halign{\hfil$\scriptstyle##$\hfil\cr<\cr\sim\cr}}}
{\vcenter{\offinterlineskip\halign{\hfil$\scriptscriptstyle##$\hfil\cr<\cr\sim\cr}}}}}
\def\ga{\mathrel{\mathchoice {\vcenter{\offinterlineskip\halign{\hfil
$\displaystyle##$\hfil\cr>\cr\sim\cr}}}
{\vcenter{\offinterlineskip\halign{\hfil$\textstyle##$\hfil\cr>\cr\sim\cr}}}
{\vcenter{\offinterlineskip\halign{\hfil$\scriptstyle##$\hfil\cr>\cr\sim\cr}}}
{\vcenter{\offinterlineskip\halign{\hfil$\scriptscriptstyle##$\hfil\cr>\cr\sim\cr}}}}}

\def \SAIT #1 #2 {{\em Mem.\ Soc.\ Astron.\ It.\/} {\bf #1}, #2}
\def \MESS #1 #2 {{\em The Messenger\/} {\bf #1}, #2}
\def \ASTRNACH #1 #2 {{\em Astron. Nach.\/} {\bf #1}, #2}
\def \AA #1 #2 {{\em Acta Astron.\/} {\bf #1}, #2}
\def \AAP #1 #2 {{\em Astron. Astrophys.\/} {\bf #1}, #2}
\def \AAL #1 #2 {{\em Astron. Astrophys. Lett.\/} {\bf #1}, L#2}
\def \AAR #1 #2 {{\em Astron. Astrophys. Rev.\/} {\bf #1}, #2}
\def \AAS #1 #2 {{\em Astron. Astrophys. Suppl. Ser.\/} {\bf #1}, #2}
\def \AJ #1 #2 {{\em Astron. J.\/} {\bf #1}, #2}
\def \ANNREV #1 #2 {{\em Ann. Rev. Astron. Astrophys.\/} {\bf #1}, #2}
\def \APJ #1 #2 {{\em Astrophys. J.\/} {\bf #1}, #2}
\def \APJL #1 #2 {{\em Astrophys. J. Lett.\/} {\bf #1}, L#2}
\def \APJS #1 #2 {{\em Astrophys. J. Suppl.\/} {\bf #1}, #2}
\def \APSS #1 #2 {{\em Astrophys. Space Sci.\/} {\bf #1}, #2}
\def \ASR #1 #2 {{\em Adv. Space Res.\/} {\bf #1}, #2}
\def \BAIC #1 #2 {{\em Bull. Astron. Inst. Czechosl.\/} {\bf #1}, #2}
\def \JSQRT #1 #2 {{\em J. Quant. Spectrosc. Radiat. Transfer\/} {\bf #1}, #2}
\def \MN #1 #2 {{\em Mon. Not. R. Astr. Soc.\/} {\bf #1}, #2}
\def \MEM #1 #2 {{\em Mem. R. Astr. Soc.\/} {\bf #1}, #2}
\def \PLR #1 #2 {{\em Phys. Lett. Rev.\/} {\bf #1}, #2}
\def \PASJ #1 #2 {{\em Publ. Astron. Soc. Japan\/} {\bf #1}, #2}
\def \PASP #1 #2 {{\em Publ. Astr. Soc. Pacific\/} {\bf #1}, #2}
\def \NAT #1 #2 {{\em Nature\/} {\bf #1}, #2}
\def \AA #1 #2 {{\em Acta Astron.\/} {\bf #1}, #2}
\def \SSREV #1 #2 {{\em Space Sci. Rev.\/} {\bf #1}, #2}
\def \aph #1 {{\em astro-ph\/} #1}
\def \IAUC #1 {{\em IAUC} #1}
\def \CJAA #1 #2 {{\em Chinese J. Astron. Astrophys.\/} {\bf #1}, #2}

\input epsf

\begin{document}

%%%%%%%%%%%%%%%%%%%%%%%%%%%%%%%%%%%%%%%%%%%%%%%%%%%%%%%%
% The title, all uppercase; if you want to split it in
% two or more lines, put a \\ macro at each line break
% example:
%   \title{TITLE: FIRST LINE\\ SECOND LINE}
%
\title{GALACTIC COLLAPSED OBJECTS}

%%%%%%%%%%%%%%%%%%%%%%%%%%%%%%%%%%%%%%%%%%%%%%%%%%%%%%%%
% The Author(S), Separated By Commas; Do not put a
% comma before the last author, use instead the \And
% macro which produces a normal ``and'' in the
% caps/small caps context
%
\author{JANUSZ ZI\'O{\L}KOWSKI}

%%%%%%%%%%%%%%%%%%%%%%%%%%%%%%%%%%%%%%%%%%%%%%%%%%%%%%%%
%
\organization{Copernicus Astronomical Center\\ul. Bartycka 18,
00-716 Warsaw, Poland}

\maketitle

\begin{abstract}
Our Galaxy  contains some $10^8 \div 10^9$ collapsed objects
(neutron stars and black holes). Our knowledge of them is based on
less than 2000 objects identified as radio pulsars, members of
X-ray binaries and magnetars. In this review I shall briefly
discuss some of their properties such as masses, magnetic fields
and rotation.

\end{abstract}
\vspace{1.0cm}

\section{Introduction}
I was asked to give a review on galactic collapsed objects (GCOs).
I should start with their definition which is rather simple: they
include neutron stars (NSs) and black holes (BHs). Both types of
objects are formed in the process of a dynamical collapse.
Collapsed objects do not include white dwarfs, which are compact
but not collapsed objects as they are not products of a collapse.
They are formed instead in the process of a very slow contraction
of the degenerate core of a star. This process takes place on a
long nuclear time scale.

As the subject of my review is very broad, I have, necessarily, to
make some arbitrary selection of the topics to be discussed. I
will start with brief discussion of the numbers of collapsed
objects in our Galaxy. Then, I will discuss the present state of
our knowledge of the masses, magnetic fields and rotation of GCOs.
These are the areas in which substantial progress was made during
the recent few years.

\section{Population of the Galactic Collapsed Objects}

According to different estimates, there are some $10^8 \div 10^9$
NSs and about one order of magnitude less (some $10^7 \div 10^8$)
BHs in our Galaxy. These estimates are based on the analysis of
the observations of X-ray Novae systems and radio pulsars on one
side, and on the stellar population synthesis calculations on the
other. Overwhelming majority of the GCOs are, for different
reasons, difficult to observe solitary objects. Inventory of the
identified GCOs contains less than 2000 objects. They include
about 320 interacting binaries (seen as X-ray binaries) which
contain about 170 recognized NSs (102 X-ray pulsars and 70 X-ray
bursters) and about 50 BH candidates (BHCs). The nature of about
100 remaining GCOs in X-ray binaries is not clearly established
yet, although in some $\sim$ 50 cases there are indications for
NSs. About 1500 GCOs are seen as radio pulsars. Majority of them
are solitary objects but about 60 are found in non-interacting
binaries. In addition, we observe about 14 young solitary NSs
which are still hot enough to emit thermal radiation from their
surfaces (most of them are also seen as radio pulsars). We observe
also about 14 magnetars (solitary NSs with extremely strong
magnetic fields: $\sim$ $10^{14} \div 10^{15}$ G). Vast majority
of our knowledge about GCOs is derived from the objects listed
above, especially from the X-ray binaries.

\section{Masses}

\subsection{Neutron stars}

Until recently, it was generally believed that  there exists
strong observational evidence indicating that all observed NSs
have (for some unclear reasons) masses close to $\sim 1.4$
M$_\odot$ ("canonical mass" of a NS). This is no longer true. The
radio pulsar  PSR J0737$-$3039B, detected recently (Lyne et al.,
2004) was found to be a companion of an earlier known millisecond
pulsar PSR J0737$-$3039A. This discovery provided us with the
first "true" binary pulsar (both components are pulsars). The
masses of two pulsars are $1.337 \pm 0.005$ M$_\odot$ and $1.250
\pm 0.005$ M$_\odot$. The second value is the lowest, precisely
determined, mass of a NS, found so far.

At the high end of the masses, the determinations are less
precise. The highest mass determined so far belongs to 4U 1700-37
and is equal $2.44 \pm 0.27$ M$_\odot$ (Clark et al. 2002) but the
neutron star nature of the compact object in this system is not
well established. The X-ray burster 4U 1820$-$30 was once claimed
to have mass $\sim 2.2$ M$_\odot$ (Zhang et al. 1998). This
estimate was based on application of the beat frequency model to
the interpretation of the kHz quasi-periodic oscillations (kHz
QPOs) observed for that burster. However, the beat frequency model
is no longer valid (seee Section 5.3.2). The best case, at
present, for high mass NS is Vela X-1. The most recent
determination (Quaintrell et al. 2003) gives for its mass $2.27
\pm 0.17$ M$_\odot$ or $1.88 \pm 0.13$ M$_\odot$ (depending on the
assumption about the inclination of the orbit and the Roche lobe
filling factor). At the 95\% confidence level, the lower limit is
$\sim 1.6$ M$_\odot$.

There is a hope that the case for a high mass NS will get stronger
in a few years time. The hope is based on a close ($P_{\rm orb} =
0.^{\rm d}26$) binary system containing the pulsar PSR J0751+1807
and a low mass ($\sim 0.19$ M$_\odot$) white dwarf. The present
estimate of the pulsar mass gives $2.1^{+0.4}_{-0.5}$ M$_\odot$
(95\% confidence, Nice et al. 2004). The present day lower limit
is therefore $\sim 1.6$ M$_\odot$, similar as for Vela X-1.
However, the precision of determinations based on pulsars timing
increases dramatically with the interval of timing. Therefore (as
opposed to the case of Vela X-1), after next few years of
observations, the mass of PSR J0751+1807 will be known with much
higher precision.

To summarize, the presently determined masses of the NSs are in
the range 1.25 M$_\odot$ to $\sim 2.0 \pm 0.4$ M$_\odot$. The
undisputed interval is only 1.25 to $\sim 1.6$ M$_\odot$ (still
substantially wider than 1.34 to 1.44 M$_\odot$ which was an
official interval for a long time).

\subsection{Black Holes}

The lowest mass galactic BHs, found so far, are: 2S 0921$-$630
($2.0 \div 4.3$ M$_\odot$, Shahbaz et al. 2004; however, the black
hole nature of the compact object in this system is not well
established) and GRO J0422+32 ($3.97 \pm 0.95$ M$_\odot$, Gelino
and Harrison 2003). On the high end of the mass scale there are
also two leaders: Cyg X-1 ($20 \pm 5$ M$_\odot$, Zi\'o{\l}kowski
2004a, 2005) and GRS 1915+105 ($21 \pm 9$ M$_\odot$, Kaiser et al.
2004).

The supermassive black hole Sgr A$^*$, residing in the center of
our Galaxy, is a special case. It is not a stellar mass BH like
the objects discussed above. However, it is a galactic compact
object and, as such, should be mentioned in our review. According
to current estimates, its mass is in the range $3 \div 4 \times
10^6$ M$_\odot$. I shall return to this object, while discussing
the spins of black holes (Section 6.2).

\section{Magnetic Fields}

\subsection{Neutron Stars}

\subsubsection{Radio Pulsars}

Applying generally accepted magnetic dipole rotator model to
explain the evolution of radio pulsars, one can easily calculate
the strength of the magnetic field of any radio pulsar from the
two very precisely measured observables: pulse period and its
derivative. From large amount of the pulsars timing data we know
that young pulsars are born with magnetic fields in the range $3
\times 10^{11}$ to $9 \times 10^{13}$ G (Manchester et al. 2004).
We know also that magnetic fields of solitary non-interacting NSs
do not decay substantially with time (at least on time scales of
up to $\sim 10^8$ years). It seems, however, that magnetic fields
of NSs experiencing accretion of matter from binary companions do
decay. The evidence comes from old radio pulsars that were spun up
during process of accretion and are seen now as, so called,
recycled pulsars (or millisecond pulsars). These recycled pulsars
have now magnetic fields in the range $8 \times 10^7$ to $2 \times
10^{10}$ G (Zi\'o{\l}kowski 1997).

\subsubsection{Accreting Neutron Stars}

Accreting NSs are seen either as X-ray pulsars (102 objects) or as
X-ray bursters (70 objects). There are also few tens of X-ray
emitters that are suspected NSs (for which neither pulses nor
bursts were detected so far). The estimates  of magnetic field are
most easy to make for X-ray pulsars. There are two ways of
estimating the field. One is based on the cyclotron absorption
lines which are now observed for some 15 X-ray pulsars.
Attributing the observed lines to electrons, one gets magnetic
fields in the range $10^{12}$ to $10^{13}$ G (Makishima et al.
1999, Grove et al. 1995). Another method is based on the
assumption that accreting NSs rotate at equilibrium periods (this
assumption is not true for slow, wind powered accretors). The
estimate using this assumption gives, for classical X-ray pulsars,
magnetic fields in the range $2 \times 10^{11}$ to $2.5 \times
10^{13}$ G (see e.g Ziolkowski 2001). The second method may be
applied also for non-pulsing accretors, if we know their spin
periods (e.g. from kHz QPOs -- see Section 4). In this way, one
gets values in the range $10^8$ to $10^9$ G for NSs (mostly X-ray
bursters) in low mass X-ray binaries (LMXBs). Similar values are
obtained for accreting millisecond pulsars, discovered recently in
a few LMXBs.

\subsubsection{Magnetars}

Magnetars (5 soft gamma repeaters and 9 anomalous X-ray pulsars)
are solitary NSs with  extremely strong magnetic fields (by about
two orders of magnitude stronger than typical young radio
pulsars). Applying classical magnetic dipole rotator model (used
for radio pulsars -- see Section 4.1.1 above) one gets magnetic
fields in the range $10^{14} \div 10^{15}$ G (see  e.g. Ziolkowski
2002). We should remember, however, that this model does not give
an accurate estimate of magnetic field for magnetars, as other
mechanisms of spin-down (e.g. strong relativistic winds) are also
present in these objects. Quite recently, an independent method
based on cyclotron lines was used to estimate the magnetic fields
of magnetars. Cyclotron lines, presumably produced by proton gas,
were observed in X-ray spectra of three magnetars: SGR 1806$-$20
($B \approx 1.0 \times 10^{15}$ G, Ibrahim et al. 2002), RXS
1708$-$4009 ($B \approx 1.6 \times 10^{15}$ G, Rea et al. 2003)
and 1E 1207$-$52 ($B \approx 1.6 \times 10^{14}$ G, De Luca et al.
2004). The last of these objects exhibits four harmonic cyclotron
lines at energies equal 0.7, 1.4, 2.1 and 2.8 keV. All these
estimates roughly agree with those based on spin period and its
derivative (assuming magnetic dipole radiation as the main
spin-down mechanism). Attributing the observed cyclotron lines to
electron gas would result in magnetic fields of the discussed
objects equal $\sim 5 \times 10^{11}$ G, $\sim 9 \times 10^{11}$ G
and $\sim 8 \times 10^{10}$ G, respectively. However, such
interpretation is very doubtful as it explains none of the well
known properties of magnetars.

One should note that some classical radio pulsars have very strong
magnetic fields ($\sim 9 \times 10^{13}$ G, McLaughlin et al.
2004) comparable to the weakest magnetic fields found in
magnetars. Yet, their other properties (X-ray emission and the
total energy output) are completely different from those of
magnetars. It seems that the strength of the magnetic field is not
the only parameter distinguishing magnetars from the rest of the
NSs population.

\subsection{Black Holes}

Practically, nothing is known about magnetic fields of galactic
BHs (but there are, of course, some theoretical speculations). We
have only indirect evidence (from some gamma ray bursts) that in
some special cases stellar mass BHs may probably possess very
powerful magnetic fields. The strength of the possible magnetic
fields around some of the observed galactic BHs, remains an
entirely open question.

\section{Rotation of Neutron Stars}

The evolution of spin of a given NS depends on the mechanism
powering its energetics (all forms of energy output). From this
point of view, we can distinguish three classes of NSs: rotation
powered (radio pulsars), magnetic field powered (magnetars) and
accretion powered (X-ray pulsars, X-ray bursters and some X-ray
emitters in LMXBs). I shall discuss only briefly the first two
classes and devote somewhat more time to the third class, as a
substantial progress was achieved recently in this area.

\subsection{Rotation Powered Neutron Stars}

Both classical radio pulsars and the recycled ones produce their
emission (mainly in the form of magnetic dipole radiation) at the
expense of their rotational energy. The observed spin periods of
classical radio pulsars are in the range from 16 ms to 8 s (the
spin correlates very well with the age -- the older the pulsar,
the longer the spin period). The recycled pulsars (spun-up in the
process of accretion of matter from the companion during their
past evolution) have the present spin periods in the range from
1.6 ms to $\sim 100$ s.

\subsection{Magnetic Field Powered Neutron Stars}

The spin periods of all magnetars are confined to a narrow range
of 5 to 12 s ($5.16 \div 11.77$). Their spin periods evolve
(increase) so fast that it is difficult to catch them at shorter
spin periods (at early phase of their evolution).

\subsection{Accretion Powered Neutron Stars}

\subsubsection{X-Ray Pulsars}

The spin periods of accreting X-ray pulsars span seven orders of
magnitude -- from 1.7 ms to $10^4$ s. Most of these periods are
probably equilibrium periods (corresponding to the balance between
spin-up accretion torque and spin-down propeller torque -- see
e.g. Zi\'o{\l}kowski 1997). The exception are slow, wind powered,
pulsars associated with supergiant companions and, possibly, some
extremely slow pulsars associated with Be stars (Zi\'o{\l}kowski
2001).

\subsubsection{Accreting Non-Pulsing Neutron Stars}

In addition to $\sim 100$ X-ray pulsars there are about 100
accreting NSs  that are not pulsators. Most of them are members of
LMXBs and are usually seen as X-ray bursters. Since long time it
was believed that these LMXBs are progenitors of recycled radio
pulsars. Therefore, the expected spin periods for NSs in these
systems were of the order of a few milliseconds. However, the
observational confirmation of these expectations appeared to be
very difficult. The breakthrough came with the discovery of
specific variability of their X-ray emission -- so callled high
frequency quasi-periodic oscillations or kHz QPOs (van der Klis et
al. 1996, van der Klis 2000). Two types of kHz QPOs were detected
in NSs sources: {\it pair QPOs} and {\it burst QPOs}.

\begin{itemize}
  \item {\bf Pair kHz QPOs}
\end{itemize}

Pair QPOs are simultaneous oscillations at two frequencies: lower
$\nu_{\rm L}$ (210 to  1050 Hz) and higher  $\nu_{\rm H}$ (500 to
1300 Hz). Both frequencies vary by a factor of up to 2 for a given
source and both increase with the increasing X-ray luminosity (or
accretion rate). However, the separation $\Delta \nu$ = $\nu_{\rm
H} - \nu_{\rm L}$ for a given source (as its X-ray luminosity
varies) remains approximately constant .

\begin{itemize}
  \item {\bf Burst kHz QPOs}
\end{itemize}

Burst QPOs are observed during Type I (thermonuclear) bursts of
some X-ray bursters (13 so far; their parameters are given in
Table 1). Their frequencies are in the range 270 to 620 Hz. These
frequencies remain constant and reproducible for a given source
(i.e. they stay the same during consecutive bursts). It was
noticed that, for a given source, the burst frequency $\nu_{\rm
B}$ was approximately equal either to $\Delta \nu$ (the separation
between two pair frequencies) or $2 \times \Delta \nu$.

\begin{table}[t!]
\begin{center}
  \caption[]{Burst QPOs Observed in Low Mass X-Ray Binaries (Chakrabarty 2004).}
  \vspace{8mm}
  \begin{tabular}{|rcl|c|c|}
\hline &&&&\\
\multicolumn{3}{|c|}{Name}&\multicolumn{1}{|c|}{$\nu_{\rm QPO}$
}&\multicolumn{1}{|c|}{Orbital Period}\\
&&&\multicolumn{1}{|c|}{[Hz]}&\\ \hline &&&&\\ 4U
1916\hspace*{-3ex}&$-$&\hspace*{-3ex}05&270&50$^{\rm m}$\\ XTE
J1814\hspace*{-3ex}&$-$&\hspace*{-3ex}338&314&4$^{\rm h}$27\\ 4U
1702\hspace*{-3ex}&$-$&\hspace*{-3ex}429&330&\\ 4U
1728\hspace*{-3ex}&$-$&\hspace*{-3ex}34&363&\\ SAX
J1808.4\hspace*{-3ex}&$-$&\hspace*{-3ex}3658&401&2$^{\rm h}$01\\
SAX J1748.9\hspace*{-3ex}&$-$&\hspace*{-3ex}2021&410&\\ KS
1731\hspace*{-3ex}&$-$&\hspace*{-3ex}260&524&\\ Aql X
\hspace*{-3ex}&$-$&\hspace*{-3ex}1&549&19$^{\rm h}$0\\ X1658
\hspace*{-3ex}&$-$&\hspace*{-3ex}298&567&7$^{\rm h}$11\\ 4U
1636\hspace*{-3ex}&$-$&\hspace*{-3ex}53&581&3$^{\rm h}$8\\ X1743
\hspace*{-3ex}&$-$&\hspace*{-3ex}29&589&\\ SAX
J1750.8\hspace*{-3ex}&$-$&\hspace*{-3ex}2900&601&\\ 4U
1608\hspace*{-3ex}&$-$&\hspace*{-3ex}52&619&\\ &&&&\\ \hline
  \end{tabular}\end{center}
\end{table}
\vspace{3mm}

\begin{itemize}
  \item {\bf Beat frequency model of kHz QPOs}
\end{itemize}

Soon  after the discovery of kHz QPOs, an interpretation known as
"beat frequency model" (BFM) was proposed (Miller et al. 1998).
This model  assumed that the higher of the pair QPO frequencies
corresponds (at least approximately) to the Keplerian angular
velocity at the inner edge of the disc (it may vary if the inner
radius of the disc changes e.g. due to variable accretion rate)
and the lower one is the beat frequency between this Keplerian
frequency and the spin of the neutron star. In this way, the
difference of the two pair QPO frequencies (which is approximately
constant for a given source) corresponds to the spin of the
neutron star. The burst QPO frequency reflects directly the spin
of the neutron star (or is equal twice the spin frequency if two
hot spots develop on the surface of a NS during the burst.

\begin{itemize}
  \item {\bf The bursting pulsars}
\end{itemize}

The next breakthrough came with the discovery of the accreting
millisecond pulsars (6 known so far, see Wijnands 2005 for the
most recent review). Their parameters are given in Table 2. Since
they are pulsars, we know unequivocally their spin periods (they
are in the range 1.67 to 5.40 msec). Two of these pulsars were
found to be also X-ray bursters exhibiting burst QPOs (compare
Table 1). These two objects provided strong evidence that
frequencies of burst QPOs reflect the true spin frequencies of the
NSs in these systems. In addition, they were found to produce also
pair QPOs! To our big surprise, we found that while in one of
these sources (SAX J1808.4$-$3658) the difference $\Delta \nu$ was
approximately equal to the spin frequency, in another (XTE
J1807-294) it was equal about half of the spin frequency! This
second case meant the end of the beat frequency model ($\Delta
\nu$ in this model might be equal twice the true spin frequency
but cannot be equal half of the spin frequency). The two bursting
pulsars confirmed the dichotomy found earlier from kHz QPOs: slow
rotators ($\nu_{\rm spin} \la 400$ Hz) have $\Delta \nu \approx
\nu_{\rm B}$, while fast rotators ($\nu_{\rm spin} \ga 400$ Hz)
have $\Delta \nu \approx  0.5 \times \nu_{\rm B}$.

\begin{table}[]
\begin{center}
  \caption[]{Millisecond X-Ray Pulsars (Wijnands 2005).}
  \vspace{8mm}
  \begin{tabular}{|rcl|c|c|}
\hline &&&&\\
\multicolumn{3}{|c|}{Name}&\multicolumn{1}{|c|}{$\nu_{\rm QPO}$
}&\multicolumn{1}{|c|}{Orbital Period}\\
&&&\multicolumn{1}{|c|}{[Hz]}&\\ \hline &&&&\\ XTE
J0929\hspace*{-3ex}&$-$&\hspace*{-3ex}314&185&43$^{\rm m}$6\\ XTE
J1807\hspace*{-3ex}&$-$&\hspace*{-3ex}294&191&41$^{\rm m}$\\ XTE
J1814\hspace*{-3ex}&$-$&\hspace*{-3ex}338&314&4$^{\rm h}$27\\ SAX
J1808.4\hspace*{-3ex}&$-$&\hspace*{-3ex}3658&401&2$^{\rm h}$01\\
XTE J1751\hspace*{-3ex}&$-$&\hspace*{-3ex}305&435&42$^{\rm m}$4\\
IGR J00291\hspace*{-3ex}&+&\hspace*{-3ex}5934&599&2$^{\rm h}$45\\
&&&&\\ \hline
  \end{tabular}\end{center}
\end{table}
\vspace{3mm}

\begin{itemize}
  \item {\bf On the rotation of NSs in LMXBs (summary)}
\end{itemize}

The accreting millisecond pulsars provided unequivocal evidence
that NSs in LMXBs do indeed rotate at millisecond periods as
expected from the analysis of recycled radio pulsars. The
presently measured periods are in the range 1.62 to 5.40 msec. The
most promising theoretical explanation of the observed kHz QPOs
seems to be the parametric epicyclic resonance theory (Abramowicz
and Klu\'zniak 2001, Lee et al. 2004). I shall describe it briefly
after discussing the spins of black holes (Section 7).

\section{Rotation of Black Holes}

\subsection{kHz QPOs}

Black Holes Candidates (BHCs) in X-ray binaries also exhibit high
frequency QPOs. These oscillations are termed kHz QPOs, similarly
as for NSs, although their frequencies are lower: 41 to  450 Hz
(see Table 3).

\begin{table}[t!]
\begin{center}
  \caption[]{High Frequency QPOs Observed in BHC Binary Systems (Remillard et al.
  2002, McClintock and Remillard 2003, Zi\'o{\l}kowski 2004b, Kaiser et al. 2004).}
  \vspace{8mm}
  \begin{tabular}{|rcl|rcl|rcl|c|}
\hline &&&&&& &&&\\
\multicolumn{3}{|c|}{Name}&\multicolumn{3}{|c|}{$\nu_{\rm QPO}$
}&\multicolumn{3}{|c|}{$M_{\rm
BH}$}&\multicolumn{1}{|c|}{comments}\\
&&&\multicolumn{3}{|c|}{[Hz]}&\multicolumn{3}{|c|}{[M$_\odot$]}&\\
\hline &&&&&&&&&\\ GRO
J1655\hspace*{-2.4ex}&$-$&\hspace*{-2.4ex}40&300\hspace*{-2.4ex}&$\pm$&\hspace*{-2.4ex}23&6.3\hspace*{-2.4ex}&$\pm$&\hspace*{-2.4ex}0.3&\\
&&&450\hspace*{-2.4ex}&$\pm$&\hspace*{-2.4ex}20&&&&\\ XTE
J1550\hspace*{-2.4ex}&$-$&\hspace*{-2.4ex}564&184\hspace*{-2.4ex}&$\pm$&\hspace*{-2.4ex}26&10.6\hspace*{-2.4ex}&$\pm$&\hspace*{-2.4ex}1.0&\\
&&&272\hspace*{-2.4ex}&$\pm$&\hspace*{-2.4ex}20&&&&\\ GRS
1915\hspace*{-2.4ex}&+&\hspace*{-2.4ex}105&41\hspace*{-2.4ex}&$\pm$&\hspace*{-2.4ex}1&21\hspace*{-2.4ex}&$\pm$&\hspace*{-2.4ex}9&\\
&&&67\hspace*{-2.4ex}&$\pm$&\hspace*{-2.4ex}5&&&&\\
&&&113\hspace*{-2.4ex}&&&&&&\\
&&&164\hspace*{-2.4ex}&$\pm$&\hspace*{-2.4ex}2&&&&\\
&&&328\hspace*{-2.4ex}&$\pm$&\hspace*{-2.4ex}4&&&&\\
&&&495\hspace*{-2.4ex}&&&&&&1.5 $\sigma$\\ 4U
1630\hspace*{-2.4ex}&$-$&\hspace*{-2.4ex}472&184\hspace*{-2.4ex}&$\pm$&\hspace*{-2.4ex}5&&&&\\
XTE
J1859\hspace*{-2.4ex}&+&\hspace*{-2.4ex}226&193\hspace*{-2.4ex}&$\pm$&\hspace*{-2.4ex}4&9\hspace*{-2.4ex}&$\pm$&\hspace*{-2.4ex}1&\\
H
1743\hspace*{-2.4ex}&$-$&\hspace*{-2.4ex}322&240\hspace*{-2.4ex}&&&&&&\\
XTE
J1650\hspace*{-2.4ex}&$-$&\hspace*{-2.4ex}500&250\hspace*{-2.4ex}&&&&&&\\
&&&&&&&&&\\ \hline
  \end{tabular}\end{center}
\vspace{3mm} {\small NOTE: 495 Hz QPO in GRS 1915+105 was detected
at only 1.5 $\sigma$ significance level}

\end{table}

The most striking feature of these QPOs is the fact, that in most
of the systems the QPO frequencies form sets of precise integral
harmonics. The fundamental frequency seems to be unique
characteristic of each black hole and, presumably, depends only on
its mass and spin. We observe 2:3 harmonics in GRO J1655$-$40 and
XTE J1550-564 and 1:2:3 harmonics in GRS 1915+105. This last
system shows, additionally, an independent set of 3:5 harmonics
(41 and 67 Hz). One can also consider the set of 113 and 164 Hz
QPOs (2:3). All theories discussing BHCs QPOs predict that the
frequencies should scale with the mass of the compact object like
$M^{-1}$. In addition, they should increase with the increasing
spin of the black hole. McClintock and Remillard (2003) found an
empirical fit using higher frequency in the 2:3 twin peak QPOs:
$\nu_3 \approx 2.8 ($M$_\odot/M)$ kHz. Abramowicz et al. (2004a)
noticed that Sgr A$^*$ (supermassive black hole in a center of
Milky Way) also satisfies this relation if one assumes that the,
recently reported, 17 minute infrared flare period corresponds to
an appropriate QPO frequency.

\subsection{Spins of BHCs (summary)}

\begin{itemize}
  \item {\bf From high frequency QPOs}
\end{itemize}

Abramowicz et al. (2004a,b) used the parametric epicyclic
resonance theory to interpret the kHz QPOs observed in three
galactic microquasars: GRO J1655$-$40, XTE J1550$-$564 and GRS
1915+105. They estimated the dimensionless angular momentum of
these BHCs to be in the range 0.7 to 0.99.

\begin{itemize}
  \item {\bf From accretion discs spectra}
\end{itemize}

Analysis of accretion discs spectra of microquasars GRO J1655$-$40
and GRS 1915+105 suggests that their dimensionless angular
momentum is in the range 0.6 to 0.9 (Zhang et al. 1997,
Gierli\'nski et al. 2001).

\begin{itemize}
  \item {\bf Sgr A$^*$}
\end{itemize}

Recently, Aschenbach et al. (2004) made use of 17 minute infrared
flare period and five different X-ray flare periodicities (100,
219, 700, 1150 and 2250 s) to estimate the mass and angular
momentum of the central Milky Way black hole. The interpretation
included Lense-Thirring precession and  epicyclic resonance
oscillations. They got 2.72 $\times 10^6 $M$_\odot$ for the mass
(somewhat less than the usual value 3.6 $\times 10^6 $M$_\odot$)
and 0.994 for the angular momentum.

\section{Parametric Epicyclic Resonance Theory}

The authors of this theory (Abramowicz and Klu\'zniak 2001,
Abramowicz et al. 2004a,b, Klu\'zniak et al. 2004, Lee et al.
2004) noticed that General Relativity (unlike Newtonian gravity)
predicts independent frequencies of epicyclical oscillation for
each spatial coordinate for a blob of matter on a perturbed orbit
around rotating compact object. Modeling done by these authors
demonstrates that there are locations at the inner accretion disc
where the coordinate epicyclical frequencies (e.g. the radial and
the azimuthal ones) form small integral ratios like 3:2, 2:1, 3:1
etc. A non-linear resonance develops at such locations leading to
the enhancement of the oscillations and producing the observed
QPOs. Modeling indicates also that, if there is a periodical
perturbing force operating in the inner disc, then the
oscillations are excited at the locations where the difference
between the coordinate frequencies is equal to the frequency of
the perturbing force or to the half of that frequency.

Parametric epicyclic resonance theory applies both to BHCs QPOs
and to the NSs QPOs. It provides a natural and elegant explanation
of the small integral ratios found for the frequencies of BHCs
QPOs. As far as NSs are concerned, the observations suggest that a
perturbing force operating at the spin frequency of the NS is
present in the inner disc. Such perturbing action of a rotating NS
is not surprising (magnetic field, surface features etc.) The
explanation of the observational facts is again natural and
elegant.

\section {Acknowledgements}

This work was partially supported by the State Committee for
Scientific Research grants No PBZ KBN 054/P03/2001 and 4 T12E 047
27.

\end{document}